\documentstyle[epsf]{npbproc}

% ...

\begin{document}
\title{QCD HADRON SPECTROSCOPY WITH STAGGERED DYNAMICAL QUARKS AT
$\beta = 5.6$\thanks{Presented by U.~M.~Heller}}
\author{K.~M.~Bitar$^a$, R.~Edwards$^a$, T.~A.~DeGrand$^b$,
        Steven~Gottlieb$^c$, U.~M.~Heller$^a$, A.~D.~Kennedy$^a$,
        J.~B.~Kogut$^d$, A.~Krasnitz$^c$, W.~Liu$^e$, M.~C.~Ogilvie$^f$,
        R.~L.~Renken$^g$, D.~K.~Sinclair$^h$, R.~L.~Sugar$^i$,
        D.~Toussaint$^j$, K.~C.~Wang$^k$}
\address{a) SCRI, The Florida State University, Tallahassee, FL 32306-4052,
USA \\
b) Department of Physics, University of Colorado, Boulder, CO 80309, USA \\
c) Department of Physics, Indiana University, Bloomington, IN 47405, USA \\
d) Department of Physics, University of Illinois, 1110 W. Green St.,
   Urbana, IL 61801, USA \\
e) Thinking Machines Corporation, Cambridge, MA 02142, USA \\
f) Department of Physics, Washington University, St.~Louis, MO 63130, USA\\
g) Department of Physics, University of Central Florida, Orlando, 
   FL 32816, USA \\
h) HEP Division, Argonne National Laboratory, 9700 S. Cass Ave., Argonne,
   IL 60439, USA \\
i) Department of Physics, University of California, Santa Barbara, 
   CA 93106, USA \\
j) Department of Physics, University of Arizona, Tucson, AZ 85721, USA \\
k) School of Physics, University of New South Wales, PO Box 1, Kensington,
   NSW 2203, Australia }

\date{12th January, 1992}

\runtitle{QCD Hadron Spectroscopy with...}
\runauthor{K.~M.~Bitar et al.} 
\volume{XXX}  
\firstpage{1} 
\lastpage{3}  

\begin{abstract}
We present preliminary results from the 1991 HEMCGC simulations with
staggered dynamical fermions on a $16^3 \times 32$ lattice at $\beta = 5.6$
with sea quark masses $am_q = 0.025$ and 0.01. The spectroscopy was done
both for staggered valence quarks with mass equal to the sea quark masses
and for Wilson valence quarks at six different values for $\kappa$, 0.1320,
0.1410, 0.1525, 0.1565, 0.1585, and 0.1600. In addition to the measurements
performed in our earlier work, we also measured the $\Delta$ and other
`extended' hadrons for staggered valence quarks and pseudo-scalar decay
constants and vector meson matrix elements, the wave function at the
origin, for Wilson valence quarks. 
\end{abstract}

\maketitle

The HEMCGC (``High Energy Monte Carlo Grand Challenge") collaboration
continued its program of spectrum calculations with two flavors of
staggered dynamical quarks in 1991, with the simulations being carried out
on the CM2 at SCRI. The spectroscopy simulation, though in principal
straightforward, has to be under control before one can trust lattice
computations of more complicated and interesting matrix elements.
As before, we used $\beta = 6/g^2 = 5.6$ and dynamical quark
masses $ma = 0.025$ and 0.01. But this time, the simulations were done on a
$16^3 \times 32$ lattice and the time direction was {\it not} doubled for
the mass measurements, as we have done previously \cite{hemcgc}. We have
also used different wall sources for the staggered spectroscopy which allowed
us to measure in addition to the nucleon, p, also the $\Delta$ as well as some
extended $\pi$'s and $\rho$'s. Here we present some preliminary
results \cite{GC91}.

We did the simulations for $ma = 0.01$ because with the lattice doubling 
we observed strange ``wiggles" in the pion effective mass \cite{hemcgc}, 
which we suspected were due to the lattice doubling in the time direction. 
For mass $ma = 0.025$, we previously used a spatial lattice of $12^3$ sites
and wanted to check for finite size effects.

For both quark masses we ran, after equilibration, 2000 trajectories of
unit length with the hybrid molecular dynamics algorithm. Staggered
spectroscopy measurements were done every $5^{\rm th}$ trajectory. We used
a conjugate gradient residue of $4 \times 10^{-5}$ and $dt = 0.02$ and 
0.01 for $ma = 0.025$ and 0.01, respectively.

The first thing to notice is that the ``wiggles" in the pion effective mass
disappeared, as shown in Fig.~1, now that we generate the configurations
already on an elongated lattice.
\begin{figure}
\epsfxsize=\columnwidth
\epsffile{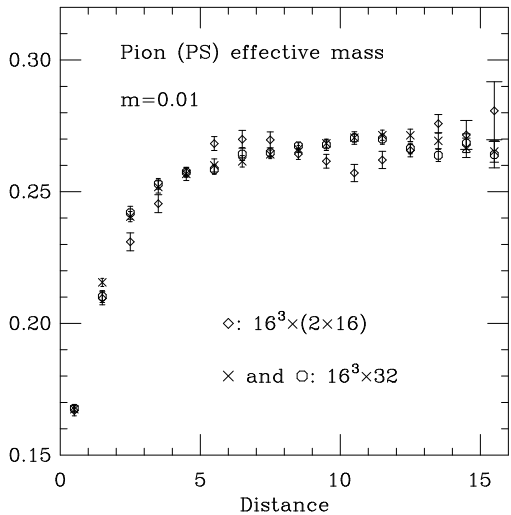}
\caption{Pion effective masses for $ma = 0.01$ on the old, doubled
($\diamond$) and the new undoubled lattices with `even-odd' source
($\times$) and `corner' source ($\circ$).}
\end{figure}

For the new spectroscopy, we used two sources, one with 1's on all even
sites in a time slice and the other with 1's on the odd sites (even-odd
source). This allows
the construction of propagators for the $\Delta$ and some extended mesons. 
The meson masses agree well with our previous results (see Table 1).
\begin{table}
\begin{tabular}{|c|c|c|c|c|}
\hline
$am$ & run & $\pi$ & $\pi_2$ & $\rho$ \\
\hline
0.01  & new & 0.270(2) & 0.350(3) & 0.516(3) \\
0.01  & old & 0.266(1) & 0.339(6) & 0.52(1)  \\
0.025 & new & 0.419(1) & 0.511(2) & 0.640(3) \\
0.025 & old & 0.415(2) & 0.499(5) & 0.63(1)  \\
\hline
\end{tabular}
\caption{Comparison of Kogut-Susskind meson masses}
\end{table}
The nucleon comes out systematically lighter. Its effective mass reaches a
plateau only at larger distances. Unfortunately, we first had a 
bug in the baryon propagators. We started reanalyzing the lattices, doing
also measurements with the old (corner)
source, with 1's in one corner of each $2^3$ cube
of a time slice, and have processed about half the $ma = 0.01$ lattices so
far. The effective nucleon masses with new and old source are shown in Fig.~2.
\begin{figure}
\epsfxsize=\columnwidth
\epsffile{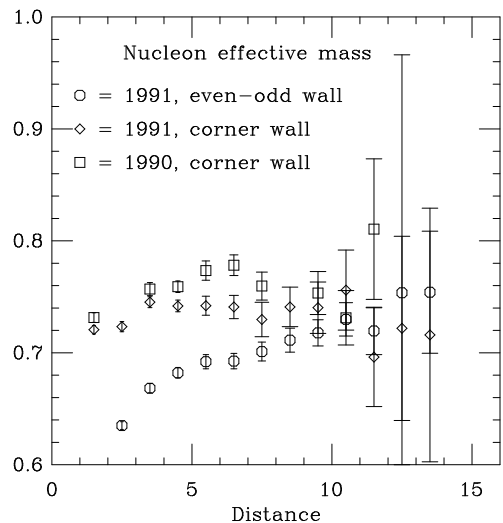}
\caption{Nucleon effective masses with new and old source and from the old
run, all with $ma = 0.01$.}
\end{figure}
It suggests that we might not have reached the asymptotic limit on our
lattices with the new (even-odd) source, and that the old (corner) source
has better overlap with the nucleon.
The new source also couples much less to the opposite parity nucleon state.
We obtain $m_p = 0.725(21)$ with the new source versus 0.742(17) with the old
source, which agrees reasonably with our earlier calculation (0.77(1)).
Because of the slow approach to the asymptotic limit of the new source
nucleon, we regard the old source to be more trustworthy.
However the new source allows measurement also of the $\Delta$ with the 
preliminary result $m_{\Delta} = 0.835(12)$ for $ma = 0.01$. The effective
mass plot of the $\Delta$ shows a quite nice plateau.

As in our previous run, we also did propagator measurements from a wall
source, every $20^{\rm th}$ trajectory, with Wilson valence quarks at six
different values for $\kappa$, 0.1320, 0.1410, 0.1525, 0.1565, 0.1585, and
0.1600. The hadron masses, for those parameter values that we had measured
in the previous set of runs, came out in satisfactory agreement (see
Table 2), except for the lightest $\Delta$.
\begin{table}
\begin{tabular}{|c|c|c|c|c|}
\hline
$\kappa$ & $\pi$ & $\rho$ & $N$ & $\Delta$ \\
\hline
0.1585 new & 0.329(3) & 0.45(1) & 0.70(1) & 0.77(2) \\
0.1585 old & 0.319(7) & 0.44(1) & 0.70(1) & 0.78(1) \\
0.1600 new & 0.218(5) & 0.39(2) & 0.61(2) & 0.63(1) \\
0.1600 old & 0.21(1)  & 0.35(1) & 0.60(1) & 0.74(1) \\
\hline
\end{tabular}
\caption{Comparison of Wilson hadron masses}
\end{table}
While we measured baryons only for equal mass quarks, we measured the meson
propagators for all pairs of  $\kappa$'s. An Edinburgh plot summarizing all
our results is shown in Fig.~3.
\begin{figure}
\epsfxsize=\columnwidth
\epsffile{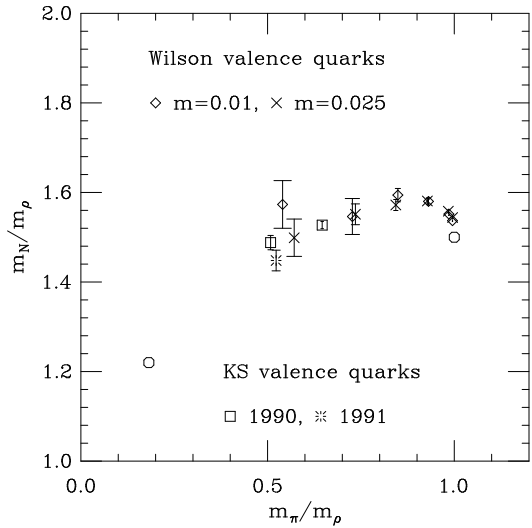}
\caption{Edinburgh plot of HEMCGC results.}
\end{figure}

We used both point and wall sinks and also measured vector and axial-vector 
current operators at the sink. This allows the extraction of certain matrix
elements, the pseudo-scalar decay constant $f_\pi$ from the axial-vector and
$1/f_\rho$, related to the wave function at the origin, from the vector
current \cite{MM}, up to $Z$ factors (except for the conserved vector
current). The determination of the $Z$ factors would require measurements
of appropriate 3-point functions, which we did not attempt (we judged our 
lattices to be too small for a meaningful measurement). Using $Z$ factors from
quenched calculations, we obtain $f_{\pi} = 105 \to 120$ \MeV\ from a local
and a nonlocal axial current, and $1/f_{\rho} = 0.21(1)$ from the conserved
vector current. These results have been extrapolated to $\kappa_c$ from the
3 largest $\kappa$'s and come out somewhat too small. For $f_{\pi}$ this
might be due to the use of $Z_A$ from quenched simulations, but for
$1/f_\rho$ this argument does not hold. Possibly we are still at too strong a
coupling, that is too far away from continuum physics.

\acknowledge{This research was supported by the U.~S.~Department of Energy
and the National Science Foundation. The computations were carried out at
SCRI, which is partially funded by DOE.}



\begin{thebibliography}{9}
\bibitem{hemcgc}
K.~M.~Bitar et al., {\sl Phys. Rev.} {\bf D42}, 3794 (1990);
K.~M.~Bitar et al., {\sl Nucl. Phys. B (Proc Suppl)} {\bf 20}, 362 (1991).
\bibitem{GC91}
K.~M.~Bitar et al., in preparation.
\bibitem{MM}
L.~Maiani and G.~Martinelli, {\sl Phys. Lett.} {\bf 178B}, 265 (1986).
\end{thebibliography}
\end{document}